\documentstyle[epsf,graphics,epsfig]{mn}

\newcommand{\etal}{{\it et~al.}}
\newcommand{\msun}{\thinspace\hbox{$M_{\odot}$}\ }

\title[The epoch of halo creation] 
{An analytic model for the epoch of halo creation}

\author[W.J. Percival \etal]{W.J.Percival,$^{1,2}$ L.Miller$^1$
	and J.A.Peacock$^2$\\
  $^1$ Dept. of Physics, University of Oxford, 
  Nuclear \& Astrophysics Laboratory, Keble Road, Oxford OX1 3RH, U.K.\\
  $^2$ Institute for Astronomy, University of Edinburgh, 
  Royal Observatory, Blackford Hill, Edinburgh EH9 3HJ, U.K.\\}

\date{Submitted for publication in MNRAS}
\begin{document}
\maketitle

\begin{abstract}
In this paper we describe the Bayesian link between the cosmological
mass function and the distribution of times at which isolated halos of
a given mass exist. By assuming that clumps of dark matter undergo
monotonic growth on the time-scales of interest, this distribution of
times is also the distribution of `creation' times of the halos. This
monotonic growth is an inevitable aspect of gravitational
instability. The spherical top-hat collapse model is used to estimate
the rate at which clumps of dark matter collapse. This gives the prior
for the creation time given no information about halo mass. Applying
Bayes' theorem then allows {\em any} mass function to be converted
into a distribution of times at which halos of a given mass are
created. This general result covers both Gaussian and non-Gaussian
models. We also demonstrate how the mass function and the creation
time distribution can be combined to give a joint density function,
and discuss the relation between the time distribution of major merger
events and the formula calculated. Finally, we determine the creation
time of halos within three N-body simulations, and compare the link
between the mass function and creation rate with the analytic theory.
\end{abstract}


\section{Introduction}  \label{sec:intro}

The hierarchical build-up of self-gravitating dark matter is thought
to drive evolution in the observable universe. The formation of clumps
of dark matter precipitates the formation of galaxies by providing a
potential well into which gas can fall and subsequently cool. Violent
mergers between equally sized halos and their associated galaxies are
thought to be important for starbursts and quasar activation. In order
to model and understand the observable universe it is therefore
essential to understand the build-up of the dark structure.

The most widely used analytic model for the distribution of mass in
isolated halos at any epoch comes from Press-Schechter (PS) theory
\cite{ps}.  By smoothing the initial field of density fluctuations on
different scales, information on the distribution of perturbation
sizes can be obtained. Linking the time at which these perturbations
collapse to the initial overdensities using the simplified spherical
top-hat collapse model allows the distribution of mass in isolated
halos at any epoch to be determined \cite{ps,peacock,bond}.

In Percival \& Miller \shortcite{ev1} (hereafter paper~I), we used the
tenets of PS theory to model the related, but distinct problem of
determining the distribution of times at which halos of a given mass
are created. Here, `creation' is defined as the epoch at which
non-linear collapse is predicted. Two derivations were given, one of
which directly used the trajectories invoked in PS theory
\cite{peacock,bond}, and one of which used Bayes' theorem to convert
from the PS mass function to a time distribution. The second
derivation required the prior for the creation time which was
calculated by examining the trajectories model.

In this paper we extend the Bayesian link between the mass function
and the creation time distribution to cover any mass function. This is
important, not only because it is known that standard PS theory is
wrong in detail (e.g. Sheth \& Tormen 1999), but especially because
the new extension applies to mass functions derived from more general
density fields including non-Gaussian models (e.g. Matarrese, Verde \&
Jimenez 2000).

First, we adopt the assumption that all clumps monotonically increase
in mass on the cosmological time scales of interest. This monotonic
growth is an inevitable aspect of gravitational instability. Every
epoch should now be thought of as a creation time for a given clump,
and we need not make the distinction between the creation time
distribution and the distribution of times at which a given halo
exists. 

In order to convert from a mass function to a distribution in time we
require the prior for the creation time. This is the rate at which
creation events occur, given no information about the halo mass. In
this work we use the spherical top-hat collapse (STHC) model to
provide a simple mechanism for determining the required rate. In
Section~\ref{sec:tophat} we derive the link between collapse time and
the overdensity at an early epoch for the STHC model within any
Friedmann cosmology. Having determined that this relation is
independent of halo mass, this leads directly to an approximation to
the prior for the creation time, described in Section~\ref{sec:time}.
This is the second major assumption adopted in this paper: that the
prior for the creation time is well approximated by this simple model
for the break-away of structure from linear expansion. This means that
following the two simple assumptions detailed above, we are able to
convert any mass function to give the distribution of epochs at which
halos of a given mass are created.

Simple models of cosmologically evolving phenomena often adopt an
important mass range rather than a specific halo mass (e.g. paper~I,
Granato \etal\ 1999). In order to use the work presented here in these
models, the joint distribution of halos in mass and creation time is
required. Although calculating the required joint probability is
formally impossible because the equations cannot be properly
normalised, a formula with the correct shape can be determined and is
presented in Section~\ref{sec:joint}.

So far we have not made a distinction between the slow accretion of
mass onto a halo and major mergers between halos. Such a distinction
is important because only major mergers are thought to play a vital
role in starbursts and quasar activation (see paper~I). The time
distribution calculated in this paper determines when halos existed
(or were created by any mechanism assuming monotonic clump growth)
which is not necessarily equal to the distribution of merger
events. This is discussed in Section~\ref{sec:mergers}.

Finally, we compare the analytic link between the mass function and
the creation rate to the results from three numerical simulations of
structure formation in different cosmological models. An analytic fit
to the mass function as described by Sheth \& Tormen \shortcite{sheth}
is adopted and is converted into a creation rate using the STHC
model. This model is compared with and shown to be in good agreement
with the numerical results.

\section{The Spherical Top-Hat Collapse Model}  \label{sec:tophat}

In this Section we analyse the STHC model which is the simplest model
for the way in which clumps of dark matter break free from linear
growth and undergo non-linear collapse. We present the derivation of
the link between the initial overdensity and the collapse time $t_{\rm
coll}$ in a form which clearly shows that this link is independent of
the mass of the overdensity. The derivation also demonstrates a method
for calculating this link within any Friedmann cosmology. Similar
derivations have been previously discussed in a variety of subsets of
this space: for an Einstein-de Sitter model, a derivation is given by
Gunn \& Gott \shortcite{gunn}, for an open $\Omega_V=0$ model by Lacey
\& Cole \shortcite{lc93}, and for a flat $\Omega_V\neq0$ universe by
Eke Cole \& Frenk \shortcite{eke}. A summary of these results is given
in Kitayama \& Suto \shortcite{kitayama2}. A numerical prescription
for the calculation of the overdensity in any cosmology has also been
developed \cite{somerville99}.

These derivations all use the same basic idea which is adopted in this
work: the behaviour of two spheres of equal mass is compared within
the cosmological framework. One of the spheres evolves with the
background density $\rho_b(t)$, while the other is perturbed by a
uniform excess density $\Delta\rho(t)$. In subsequent analysis, a
subscript `$b$' denotes that a quantity relates to the sphere with
background density, and `$p$' to the perturbation.

Matter is assumed to be an ideal fluid with no pressure and the
universe is modelled as spherically symmetric around the
perturbation. As a consequence of Birkhoff's theorem, the
gravitational field of both the perturbation and the background is
described by a Robertson-Walker (RW) metric with curvature constant
$K$, and RW scale factor $a(t)$. The behaviour of such perturbations
is governed by Friedmann's equation which we will consider in the
form:
\begin{equation}
  \left(\frac{da}{dt}\right)^2+K=\frac{2GM}{a}+(H_0^2\Omega_V)a^2
  \label{eq:friedmann1}
\end{equation}
where $M$ is the mass inside the sphere. Note that in order to compare
spheres with different behaviour, we do {\em not} normalise the scale
factor $a(t)$ to equal the curvature scale (by dividing by
$\sqrt{|K|}$) so $K$ is allowed to take any real value.

To calculate the behaviour of the overdensity at an early time, we
note that a series solution for $a(t)$ in the limit $t\to0$ can be
obtained for Equation~\ref{eq:friedmann1}. This is given by $a=\alpha
t^{2/3}+\beta t^{4/3}+O(t^{6/3})$, where:
\begin{equation}
  \alpha=\left(\frac{9GM}{2}\right)^{1/3}, \hspace{1cm}
  \beta=\frac{3K}{20}\left(\frac{6}{GM}\right)^{1/3}.
  \label{eq:ab}
\end{equation}
Using the fact that the spheres contain equal mass, the behaviour of
$\delta(t)\equiv\Delta\rho(t)/\rho_b(t)$ in the limit $t\to0$ is given
by:
\begin{eqnarray}
  \lim_{t\to0}\left(\frac{\Delta\rho(t)}{\rho_b(t)}\right)
    &=&\lim_{t\to0}\left(\frac{a_b(t)^3}{a_p(t)^3}-1\right) \nonumber \\
    &=&\frac{3}{\alpha}(\beta_p-\beta_b)t^{2/3}+O(t^{4/3}). 
  \label{eq:limits}
\end{eqnarray}
Defining: 
\begin{equation}
  \epsilon=\frac{K}{(GMH_0)^{2/3}}, 
\end{equation}
the present day normalisation of Equation~\ref{eq:friedmann1} gives
that for a sphere of uniform background density:
\begin{equation}
  \epsilon_b=(\Omega_M+\Omega_V-1)
    \left(\frac{2}{\Omega_M}\right)^{2/3},
  \label{eq:kbg}
\end{equation}
We can now combine Equations~\ref{eq:ab},~\ref{eq:limits}~\&~\ref{eq:kbg} 
to determine the behaviour of $\delta(t)$ in the limit $t\to0$ as a 
function of $\epsilon_p$:
\begin{displaymath}
  \lim_{t\to0}\delta(t)=
    \frac{9}{20}\left(\frac{4}{3}\right)^{1/3}
\end{displaymath}
\begin{displaymath}
  \hspace{1cm}
    \times
    \left[(\Omega_M+\Omega_V-1)
    \left(\frac{2}{\Omega_M}\right)^{2/3}-\epsilon_p\right](H_0t)^{2/3} 
\end{displaymath}
\begin{equation}
  \hspace{1cm}
    +O[(H_0t)^{4/3}]
  \label{eq:ktolimd}
\end{equation}

If the field of perturbations is linearly extrapolated to present day
and normalised here, the approximation of Carroll, Press \& Turner
\shortcite{carroll} for the ratio of the current linear amplitude to
the Einstein-de Sitter model can be used to extrapolate the limiting
behaviour of $\delta$ to this epoch. The extrapolated limit,
$\delta_{\rm lim}$, is related to $\epsilon_p$ by:
\begin{displaymath}
  \delta_{\rm lim} \simeq \frac{3}{8}\left(4\Omega_M\right)^{2/3}
    \left[(\Omega_M+\Omega_V-1)
      \left(\frac{2}{\Omega_M}\right)^{2/3}
      -\epsilon_p\right]
\end{displaymath}
\begin{equation}
  \hspace{3mm}
  \times
    \left[\Omega^{4/7}_M-\Omega_V+
    \left(1+\frac{1}{2}\Omega_M\right)
    \left(1+\frac{1}{70}\Omega_V\right)\right]^{-1}.
  \label{eq:ktodc}
\end{equation}
A similar formula is possible if the field of fluctuations is
normalised at any other epoch. Note that $\delta_{\rm
lim}\propto(\epsilon_p+{\rm constant})$ and the time dependence of
$\delta_{\rm lim}$ is given by that of $\epsilon_p$.

For the perturbation, the radius of maximum expansion can be
calculated from Equation~\ref{eq:friedmann1}: this radius corresponds
to the first positive root of the equation $2GM+H_0^2\Omega_Va^3-Ka=0$
denoted by $a_{\rm max}$. This leads to a necessary and sufficient
condition for the perturbation to collapse: that such a (finite) root
exists. Because of the symmetry in Equation~\ref{eq:friedmann1}, this
model predicts that the perturbation will collapse to a singularity at
a time equal to twice the time required to reach maximal expansion:
\begin{equation}
  H_0t_{\rm coll}=2\int^{a^*_{\rm max}}_{0}
    \left(\frac{2}{a^*}+\Omega_V(a^*)^2
    -\epsilon_p\right)^{-1/2}\,da^*
  \label{eq:ttok}
\end{equation}
where we have changed from $a$ to $a^*=aH_0^{2/3}/(GM)^{1/3}$, and
$a^*_{\rm max}$ is the first positive root of the equation:
\begin{equation}
  2+\Omega_V(a^*)^3-\epsilon_p a^*=0.
  \label{eq:amax}
\end{equation}
Although collapse to a singularity does not occur in practice, the
virialisation epoch is assumed to be similar to $t_{\rm coll}$.  

For perturbations that collapse, $\delta_{\rm lim}$ is called the
`critical' density and is denoted $\delta_c$. Equation~\ref{eq:ktodc}
then gives $\delta_c(\epsilon_p)$.
Equations~\ref{eq:ttok}~\&~\ref{eq:amax} give $t_{\rm
coll}(\epsilon_p)$, and the combination of these three Equations gives
the required link between $\delta_c$ and the collapse time. Note that
these Equations are independent of the perturbation mass, and
therefore so is the link between the initial overdensity and the
collapse time.

In practice we wish to use these Equations to calculate
$\delta_c(z_{\rm coll},\Omega_M,\Omega_V)$ or $d\delta_c(z_{\rm
coll},\Omega_M,\Omega_V)/dt$ where $z_{\rm coll}$ is the collapse
redshift. Unfortunately this is not easy as
Equations~\ref{eq:ttok}~\&~\ref{eq:amax} cannot be inverted to give
$\epsilon_p(t_{\rm coll})$. The procedure adopted is as follows: the
collapse time can be numerically determined from $z_{\rm coll}$ using
the Friedmann equation for the background cosmology. $\epsilon_p$ can
be determined numerically using
Equations~\ref{eq:ttok}~\&~\ref{eq:amax}, and $\delta_c$ can be
calculated using Equation~\ref{eq:ktodc}. $d\delta_c/dt$ can be
calculated numerically from $\delta_c(t_{\rm coll})$ and is discussed
further in the next Section.

For the subset of cosmological models with $\Omega_V=0$, the above
procedure is simplified and analytic formula can be obtained for
$\delta_c$. In this case, Equation~\ref{eq:ttok} reduces to:
\begin{equation}
  H_0t_{\rm coll}=2\int^{2/\epsilon_p}_{0}
    \left(\frac{2}{a^*}-\epsilon_p\right)^{-1/2}\,da^*.
\end{equation}
Making the substitution $\tan(\theta)=(2/a^*-\epsilon_p)^{-1/2}$, this
integral can be solved to give:
\begin{equation}
  H_0t_{\rm coll}=\frac{2\pi}{\epsilon_p^{3/2}}.
  \label{eq:nocstetot}  
\end{equation}

We now show that these Equations provide the result of Gunn \& Gott
\shortcite{gunn} for an Einstein-de Sitter cosmology. In this case,
substituting Equation~\ref{eq:nocstetot} into Equation~\ref{eq:ktodc}
gives that:
\begin{equation}
  \delta_c(t_{\rm coll}) = \frac{3}{20}
    \left(\frac{8\pi}{H_0t_{\rm coll}}\right)^{2/3}.
\end{equation}
We can now change from collapse time to collapse redshift to give:
\begin{eqnarray}
  \delta_c(z_{\rm coll})&=&\frac{3}{20}(12\pi)^{2/3}(1+z_{\rm coll}) 
    \nonumber \\
    &\simeq&1.69(1+z_{\rm coll}),
\end{eqnarray}
which is the equation of Gunn \& Gott \shortcite{gunn}.

\section{From a mass function to a time distribution}  \label{sec:time}

In this Section we show how to convert from a mass function to the
distribution of times at which isolated halos of a given mass
exist. First, we make the assumption that the mass of any clump is a
monotonically increasing function of time so that the mass will
increase between any two epochs. This is true for Press-Schechter
theory (see paper~I). Note that this mass growth is not constrained to
be continuous and the mass is allowed to undergo instantaneous finite
increases, or `mass jumps'. Following this assumption, every epoch at
which a halo exists should also be considered as a `creation' epoch:
every halo is a new isolated halo of some mass. The distribution of
`creation events' is therefore the same as the distribution of times
at which the halos exist. Note that by definition this only applies to
isolated halos which have not been subsumed into larger objects.

In this paper we have called this epoch the `creation' time of a halo
in order to avoid confusion with other authors definitions of the
`formation' time of a halo. Note that this semantic change was not
adopted in paper~I. The `formation' time of a halo was defined by
Lacey \& Cole \shortcite{lc93} as the latest time when the largest
progenitor of a halo has a mass less than half that of the final
halo. This definition makes sense if we are discussing a non-evolving
quantity, say the existence of a galaxy halo, and wish to know when it
was formed given that it exists at present day. However, suppose we do
not know anything about the build-up of a halo before or after it has
mass $M$ and only wish to know when it was likely to have
existed. This Lacey \& Cole definition of `formation' cannot help us
for we do not know the time and mass from which to determine
progenitors: progenitors of what?

In order to calculate the probability density function (pdf) of the
times at which halos exist, we consider the set of all possible times
and all possible halo masses. This is the `sample space' of our
`experiment'. The experiment consists of choosing a particle, or small
mass element, and an `event' is given by any subset of the sample
space: for instance that the particle is part of a halo of mass
$M_1<M<M_2$ created at time $t_1<t<t_2$, or that the particle inhabits
a halo of mass $M$, created at time $t$.

Denoting a generic pdf by the function $f$, the mass function is given
by $f(M|t)\,dM$, the distribution of halo masses at a given
epoch. This is equal to $Mn(M)/\rho$ where $n(M)$ is the number
density of halos. The pdf we wish to determine is given by
$f(t|M)\,dt$, the distribution of times at which halos of mass $M$
were created. Note that our assumption of monotonic mass growth means
that $t$ is the same variable in $f(M|t)\,dM$ {\em and}
$f(t|M)\,dt$. These pdfs are then related by the following formula,
based on Bayes' theorem:
\begin{equation}
  f(t|M)\,dt=\frac{f(M|t)\,dM\,f(t)\,dt}
    {\int_0^{\infty}\left[f(M|t)\,dM\,f(t)\right]\,dt},
  \label{eq:bayes}
\end{equation}
where $f(t)\,dt$ is the normalised prior for time, or the distribution
of creation events in time given no information about the mass of
halo. In paper~I we calculated the prior using the Brownian random
walks invoked in PS theory with a sharp $k$-space filter. In order not
to bias the distribution of up-crossings within this model, we assumed
a uniform prior for $\delta_c$. 

The reason the prior is uniform in $\delta_c$ follows from the STHC
model. Within this model, it is the density $\delta$ associated with a
particle that is important, and the barrier has to move from $\delta$
to $\delta-d\delta$ for `halo creation' to have occurred. Given that
the mass of all clumps monotonically increases, all particles will be
associated with creation events at any $\delta_c$. Following these two
observations, any two equal width intervals in $\delta_c$ should
contain equal `numbers' of halo creation events.

The derivation presented in the previous Section showed that for the
STHC model, the link between the critical overdensity and the collapse
time is independent of the perturbation mass. Therefore, given no
information about the mass contained within a perturbation, the pdf
for the time at which the perturbation collapses should be assumed to
be proportional to the time derivative of $\delta_c(t)$.  This gives
the rate at which the collapse threshold $\delta_c(t)$ crosses the
initial overdensities. This can be calculated numerically from the
following formula:
\begin{displaymath}
  \frac{d\delta_c}{dt}\propto\frac{d\epsilon_p}{dt}=
\end{displaymath}
\begin{equation}
  \left[\frac{d}{d\epsilon_p}
    \left(\int\limits_0^{a^*_{\rm max}(\epsilon_p)}
    \left(\frac{2}{a^*}+\Omega_V(a^*)^2-\epsilon_p\right)
    ^{-\frac{1}{2}}da^*\right)\right]^{-1},
\end{equation}
where $a^*_{\rm max}$ is the first positive root of the equation
$2+\Omega_V(a^*)^3-\epsilon_p a^*=0$. Note that for
cosmologies with $\Omega_V=0$, the above equation can be
analytically solved as for Equation~\ref{eq:nocstetot}, and the
derivative $d\delta/dt$ is proportional to $t^{-5/3}$.

Unfortunately, $d\delta_c(t)/dt$ cannot be normalised so that it
integrates over all time to give unity. This means that we cannot
simply take a multiple of $d\delta_c(t)/dt$ as the prior for the
collapse time. However, we can still use Equation~\ref{eq:bayes} by
making use of a mathematical trick and placing an arbitrary upper
limit on $t$, $t_u$, which can be removed later without affecting the
result. This gives that:
\begin{equation}
  f(t|M)\,dt=\lim_{t_u\to\infty}\left[\frac{f(M|t)\,dM\,f(t,t_u)\,dt}
    {\int_0^{t_u}\left[f(M|t)\,dM\,f(t,t_u)\right]\,dt}\right]
  \label{eq:bayes2}
\end{equation}

The connection between the mass function and the creation rate of
halos presented in this Section is consistent with that of paper~I:
the prior for time used is exactly the same. We have merely shown that
adopting the STHC model for the rate at which structures are created
allows {\em any} mass function to be converted to give the pdf of the
time at which a halo of a given mass is created.

\section{The relation with the multiplicity function}

Changing variables from mass to a function of
\begin{equation}
  \nu\equiv\frac{\delta_c}{\sigma_M}
\end{equation}
alters the form of the standard PS mass function to one which is
invariant with respect to time. Here $\sigma_M$ is the rms fluctuation
of the initial density field smoothed with a top-hat filter on a scale
related to mass $M$. Unless stated otherwise we change variables in
the mass function from M to $\ln\nu(M,t)$. The normalised pdf
$f(\ln\nu|t)$ is called the multiplicity function and is related to
the mass function by:
\begin{equation}
  f(M|t)=Af(\ln\nu|t)\left.\frac{\partial\ln\nu}{\partial M}\right|_t,
  \label{eq:multiplicity_M}
\end{equation}
where $A$ is a normalisation constant. Note that we have retained the
condition on time in $f(\ln\nu|t)$, to emphasise that we are still
concerned with the distribution of halos at a particular
epoch. Although the multiplicity function has a form which is
invariant with respect to time, it still gives the distribution of
$\ln\nu$ we would expect for halos given a particular time. This is
{\em not} the same as the the distribution of $\ln\nu$ we would obtain
if we chose halos at random in both mass and time, or the distribution
of $\ln\nu$ we would obtain if we chose halos only of a particular
mass.

If the mass function can be written in a form which is independent of
time as described above, then under the same change of variables, the
creation rate becomes independent of halo mass. The resulting pdf
$f(\ln\nu|M)$ is now only valid if we are examining the distribution
of halos at fixed mass. Following the notation adopted above, this is
given by:
\begin{equation}
  f(t|M)=Af(\ln\nu|M)\left.\frac{\partial\ln\nu}{\partial t}\right|_M.
  \label{eq:multiplicity_t}
\end{equation}

\section{The joint distribution of halos in mass and time}  \label{sec:joint}

Because the mass of each halo is assumed to monotonically increase
with time, within any interval of mass and time, an infinite number of
`creation events' occur. This means that the joint probability of the
existence of a halo in {\em both} mass and time cannot be properly
normalised.

Equation~\ref{eq:bayes2} gives the link between two pdfs, the mass
function and the creation rate using a mathematical trick to cope with
an un-normalised prior in time. The numerator of this equation is the
joint distribution of halos in mass and time,
$f(M,t)\,dM\,dt=f(M|t)\,dM\,f(t)\,dt$. The denominator is not a
function of time: it only normalises the resulting formula so
$f(t|M)\,dt$ integrates to unity. Following this argument, given a
mass function, multiplying by $d\delta_c(t)/dt$ creates a function
with {\em both} the correct mass and time behaviour. This joint
distribution function (not a pdf) has the same mass dependence as
$f(M|t)$ and the same time dependence as $f(t|M)$.

As an example we consider the fitting function of Sheth \& Tormen
\shortcite{sheth} to the multiplicity function determined from the
results of N-body simulations for different cosmological parameters:
\begin{displaymath}
  \frac{Mn(M)}{\rho}\,dM=f(M|t)\,dM=f(\ln\nu|t)\,d\ln\nu 
\end{displaymath}
\begin{equation}
  \hspace{10mm}
  = A\sqrt{\frac{2}{\pi}}\left(1+\frac{1}{\nu'^{2p}}\right)\nu'e^{-\nu'^2/2}
    \,d\ln\nu,
  \label{eq:sheth}
\end{equation}
where $\nu'=a^{1/2}\nu$ and $a$~\&~$p$ are parameters. Note that Sheth
\& Tormen displayed this formula using a different notation to that
adopted here, although parameters $a$ and $p$ are the same in both
cases. $A$ is determined by requiring that the integral of
$f(\ln\nu|t)$ over all $\ln\nu$ gives unity. Sheth \& Tormen found
best fit parameters $a=0.707$ and $p=0.3$ for their simulations and
group finding algorithm. The standard PS multiplicity function has
$a=1$, $p=0$ and $A=1/2$. Unless stated otherwise, by standard PS
theory, we refer to the adoption of this multiplicity function
combined with top-hat filtering (to calculate $\sigma_M^2$). In order
to convert this function to provide a model of both the time and mass
of halo creation events, all we need to do is to multiply by
$d\delta_c/dt$.

For standard PS theory, writing $\nu$ explicitly in terms of
$\sigma_M^2$ and $\delta_c$ we find that the joint distribution of the
existence of a halo in mass and time reduces to:
\begin{displaymath}
  f(M,t)\,dM\,dt= \frac{\delta_c}
    {(2\pi)^{1/2}\sigma_{M}^{3}}
\end{displaymath}
\begin{equation}
  \hspace{10mm}
  \times
    \exp\left(-\frac{\delta_c^{2}}{2\sigma_M^2}\right)
    \left|\frac{d\sigma_M^2}{dM}\right|
    \left|\frac{d\delta_c}{dt}\right|\,dM\,dt, 
  \label{eq:fmnt}
\end{equation}
Although not normalised, such a formula integrated over any two areas
of the mass-time plane will provide the correct relative number
densities.

Note that this is {\em not} the same formula as obtained by simply
multiplying the mass function with the creation time distribution at
fixed mass. This would be inconsistent within a Bayesian framework and
would produce a joint density function which lacks the correct mass
and time behaviour: the form of each conditional pdf is altered by the
other. Care should therefore be taken when using the creation rate in
models which also include the mass function.

\section{The relation with merger events}  \label{sec:mergers}

So far, we have only been concerned with the epoch at which a halo is
created. However, there is an important distinction between major
mergers and the slow accretion of mass when applying the results in
models of certain cosmological phenomena. For instance, only violent
merger events are thought to be important for starbursts and quasar
activation. In paper~I, we showed that for standard PS theory with a
sharp $k$-space filter, if mass jumps in a particular trajectory
correspond to merger events, then the distribution of mergers is the
same as that of the build-up of matter from all types of creation
event. This is because the trajectories are Brownian random walks
which have the special property that their form is independent of the
initial point.

Given only the mass function and the assumptions outlined above it is
not possible to determine how each clump increases in mass, only the
distribution of times at which it reaches a certain mass. More
information about the build-up of individual clumps is required before
the distribution of major mergers can be determined. Such information
is available in PS theory and follows from the argument that each
trajectory gives the history of the halo masses in which a particular
small mass element resides.

\section{Description of the Numerical Simulations}  \label{sec:simulations}

A direct approach to modelling structure formation is to simulate the
evolution of the mass density of the Universe using a distribution of
softened particles. We have run three such simulations using the Hydra
N-body, hydrodynamics code \cite{couchman} with $128^3$ dark matter
particles to model the build-up of halos for three different
cosmological models, described in Table~\ref{tab:cosmo}.

\begin{table}
  \centering
  \begin{tabular}{cccccc} 
  model & $\Omega_{M}$ & $\Omega_V$ & $\Gamma$ &
    $\sigma_{8}$ & h \\ \hline 	
  $\Gamma$CDM  & 1   & 0   & 0.25 & 0.64 & 0.5 \\
  OCDM         & 0.3 & 0   & 0.15 & 0.85 & 0.5 \\
  $\Lambda$CDM & 0.3 & 0.7 & 0.15 & 0.85 & 0.5 \\
  \end{tabular}
  \caption{Table showing the parameters of the different cosmological
  models adopted in the three N-body simulations.}  \label{tab:cosmo}
\end{table}

In order to determine the {\em rate} at which halos are created within
these simulations, we output particle positions at a large number of
times. For the $\Gamma$CDM simulation, we output particle positions at
362 different epochs, separated by approximately equal intervals in
time. For the OCDM simulation the number of outputs was 345 and for
the $\Lambda$CDM simulation, 499. The box size chosen was
100\,$h^{-1}$Mpc for all three simulations which gave a particle mass
of $2.6\times10^{11}$\,\msun for $\Gamma$CDM and
$7.9\times10^{10}$\,\msun for the other two simulations. Groups of
particles were found for each output using a standard
friends-of-friends algorithm with linking length set to $b=0.2$ times
the mean interparticle separation.

\section{Fitting to the Mass Function}

\begin{figure}
  \setlength{\epsfysize}{13.4cm}
  \centerline{\epsfbox{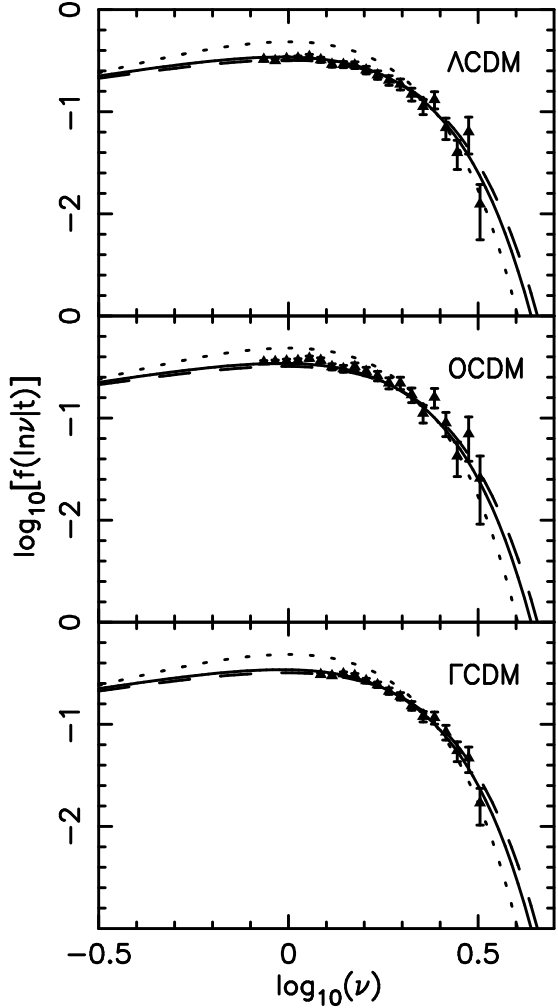}} \caption{The measured
  multiplicity function from each of the three simulations analysed
  for all groups containing over 45 particles (solid triangles plotted
  with Poisson error bars). See Section~\ref{sec:simulations} for
  details of these simulations. In compiling these data in this way we
  have assumed that the transfer of variables from $M$ to $\ln\nu$
  where $\nu\equiv\delta_c/\sigma_M$ does make the distribution of
  masses independent of epoch as found by Sheth \& Tormen
  \protect\shortcite{sheth}. For comparison we have also plotted the
  best fit model to their data (dashed line) and the predictions of
  standard PS theory (dotted line). The solid line shows the best fit
  model to our data allowing the parameters of the Sheth \& Tormen
  \protect\shortcite{sheth} fitting function to vary.}
\label{fig:nbody_mult}
\end{figure}

The multiplicity function averaged over all output times is presented
from each of the simulations in Fig.~\ref{fig:nbody_mult}. Here we
have only considered groups containing over 45 particles in order to
limit the number of false detections due to numerical effects. In
compiling the data in this way, we have assumed that converting from
mass to $\ln\nu$ does indeed convert the form of the mass function
into one which is independent of epoch. This Figure has been produced
in such a way as to be directly comparable with figure~2 of Sheth \&
Tormen \shortcite{sheth}. For comparison we also plot their best fit
model and the predictions of standard PS theory.

We have also plotted the model of Sheth \& Tormen \shortcite{sheth}
(Equation~\ref{eq:sheth}) after allowing the parameters to vary to
simultaneously fit the data from all three simulations. We find
slightly different best fit parameters to those of Sheth \&
Tormen. Our best fit parameters are $a=0.774, p=0.274$, compared to
standard PS theory $a=1, p=0$ and Sheth \& Tormen $a=0.707,
p=0.3$. Note that the difference between our best fit values and those
of Sheth \& Tormen could be explained by the different group finding
algorithms used.

\section{Comparison between the Analytic and Numerical Halo Creation Rates}

Although we have argued that the monotonic increase in mass means that
all epochs are `creation' times for a given halo, we cannot simply
compare the creation rate formulae with the distribution of halo
numbers at different epochs: each halo should only be counted once. To
determine the distribution of creation times of halos of mass $M$, we
therefore sequentially analysed the FOF output from $z=50$ to present
day. All halos of mass $>M$ were examined at each epoch to determine
whether they were `new'. The definition of `new' adopted was that at
least half of the particles in a halo were not included in any halo of
mass $>M$ at a previous output time. The number of these halos in the
required mass range was taken to be the minimum number which could
have been created between that output time and the previous one. In
order not to miss creation events where a halo was created and
subsumed into a larger halo all within the time interval between two
outputs, we analysed the progenitors of all new halos with mass
greater than the required range. Those with a progenitor distribution
at the previous step which could sum to a halo of the required mass
were recorded as a possible halo of the required mass. In this way we
determined the minimum and maximum mass which could have been created
in each time interval between output from the simulation.

\begin{figure*}
  \setlength{\epsfysize}{13.2cm}
  \centerline{\epsfbox{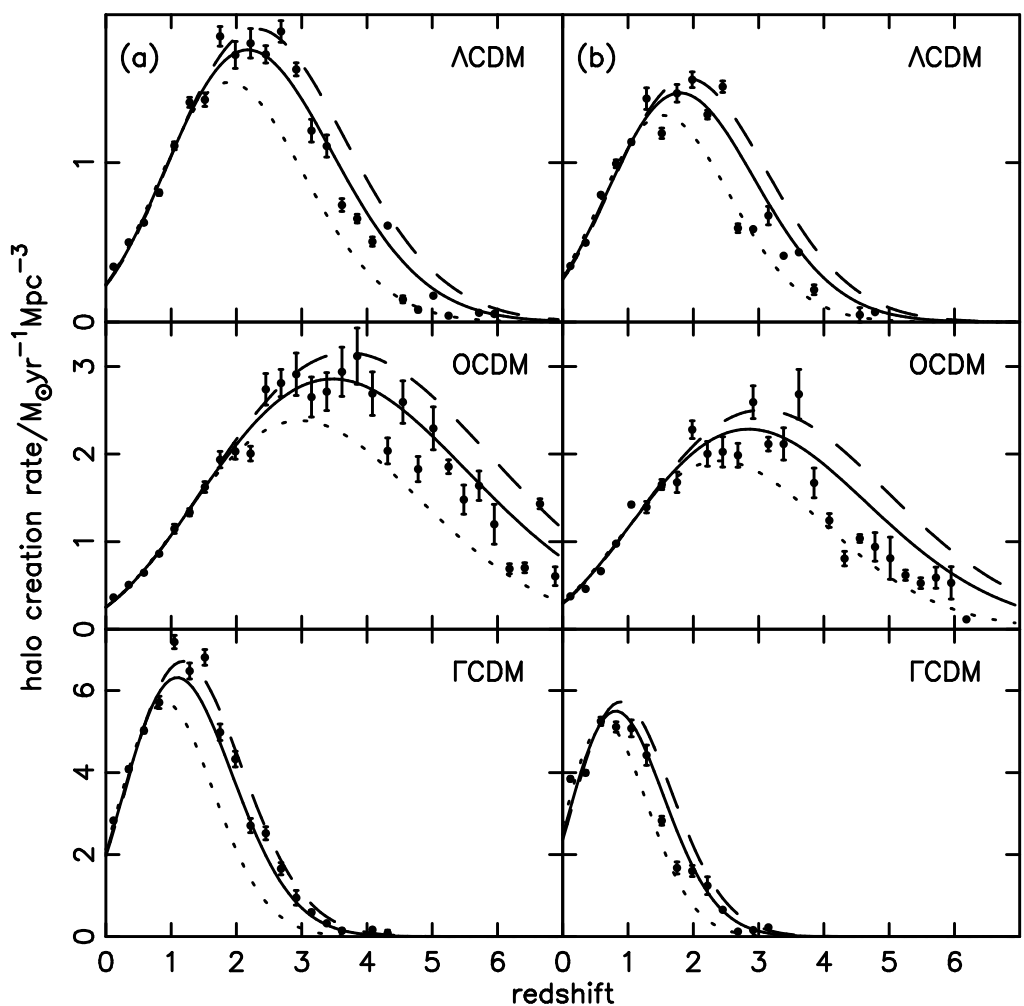}} \caption{The creation rate
  of halos as determined from the three cosmological simulations
  (solid circles) for a) 45-50 particles in a halo and b) 100-110
  particles. These group sizes correspond to masses of aproximately a)
  $1.2\times10^{13}\msun$ for $\Gamma$CDM and $3.6\times10^{12}\msun$
  for the two other cosmologies, and b) $2.6\times10^{13}\msun$ for
  $\Gamma$CDM and $7.9\times10^{12}\msun$ for the two other
  cosmologies. The error bars show the binned minimum and maximum mass
  which could have been created in halos of the required masses in
  each time interval. Symbols are plotted half way between the
  two. Note that the errors bars do not include counting errors. For
  comparison we also plot the three models of the mass function
  multiplied by $d\delta_c/dt$ with lines as in
  Fig.~\ref{fig:nbody_mult}. The models have been normalised to the
  low redshift data.}
\label{fig:nbody_form}
\end{figure*}

In Fig.~\ref{fig:nbody_form} we plot the creation rate for halos
within two narrow mass ranges. In order to obtain the maximum number
of creation events, we have used relative low numbers of particles in
each group. Data are plotted for groups of between $45-50$ and
$100-110$ particles. These distributions are compared with the three
multiplicity functions plotted in Fig.~\ref{fig:nbody_form}, converted
into creation rates by multiplying by $d\delta/dt$ for halos of mass
equivalent to 45 or 100 particles. These curves have been normalised
to the low redshift data.

All of the models reproduce the decrease in creation events to present
day seen in the simulations. As output from the simulation occured
after approximately equal intervals of time, the high redshift data
suffers as the intervals contain relatively more creation events. This
means that we cannot precisely follow the build-up of the clumps, and
the difference between the maximum and minimum mass which could have
been created in each bin is increased. This is particularly noticable
in the OCDM simulation where halos are created at earlier times and we
have fewer outputs from the simulation.

However, there is evidence that the solid line (calculated from the
best fit to the mass function) also fits the creation rate data the
best out of the three models plotted. As a rough guide to this, the
root mean square value between the plotted data points and the model
is 3.4 for this curve, compared to 6.7 for standard PS theory, and 5.0
for the best fit model of Sheth \& Tormen \shortcite{sheth}. Note that
the form of the creation rate is strongly dependent on the parameter
$a$ in Equation~\ref{eq:sheth}, and only weakly dependent on parameter
$p$. This is consistent with the importance of these parameters for
the mass function: parameter $a$ controls the position of the
high-mass cut-off, whereas parameter $p$ controls the low-mass tail of
the distribution.

\section{Conclusions}

We have demonstrated a simple method for linking any mass function to
the corresponding distribution of times at which isolated halos of a
given mass are created. In order to provide this link we adopted the
assumption that the time scales of interest are those over which the
mass of every clump can be thought of as monotonically increasing. The
prior for the collapse time was estimated using the STHC model which
ties in directly with PS theory, although the method does not use any
of PS theory beyond that of the STHC model. We have presented a new
derivation of the link between the collapse time and initial
overdensity for this model which explicitly shows that this link is
independent of the halo mass and is applicable in any Friedmann
cosmology. Multiplying the mass function by a function with no mass
dependence and proportional to the time derivative of the critical
overdensity then provides a joint density function with the correct
behaviour for the creation of a halo in mass {\em and}
time. Integrating over the resulting joint density function will give
the correct relative number densities of halos within different mass
and time intervals.

We have extended the analysis of N-body simulation results presented
in paper~I to cover three simulations of the build-up of dark matter
within different cosmological models. Rather than using PS theory, we
have demonstrated how a fit to the mass function may be converted to
give a creation rate. Out of the three functions we have compared to
the mass function data, the best fit model for these data when
converted to a creation rate also fits the creation rate data the
best. This gives us confidence that the formalism presented here is
sound, and should give accurate results in more general situations, in
particular non-Gaussian models.

\section{Acknowledgements}
We are grateful for the use of the Hydra N-body code \cite{couchman}
kindly provided by the Hydra consortium.

\end{document}